\documentclass[12pt,epsf,qsymbols]{article}
\pdfoutput=1
\usepackage[utf8]{inputenc}
\usepackage[normalem]{ulem}
\usepackage[english]{babel}
\usepackage{tabularx}
\usepackage{array}
\usepackage{graphics}
\usepackage[pdftex]{graphicx}
\usepackage{psfrag}
\usepackage{epsfig}
\usepackage{subfig}
\usepackage{amsmath}
\usepackage{mathtools}
\usepackage{amssymb}
\usepackage{bbold}
\usepackage{ulem}
\usepackage{setspace}
\usepackage{rotating}
\usepackage{colortbl}
\usepackage{longtable}
\usepackage{slashed}
\usepackage{braket}
\usepackage{lineno}
\makeatletter
\usepackage{textcomp}
\usepackage[usenames,dvipsnames]{xcolor}
\usepackage{relsize}
\usepackage{cite}

\usepackage{float}

\usepackage{bbm}
\usepackage{bm}
\usepackage{enumerate}
\usepackage{amsmath}
\usepackage{verbatim}

\usepackage{fancyhdr}

\usepackage{multirow}
\usepackage{arydshln}
\usepackage{enumitem}

\usepackage{hyperref}
\hypersetup{colorlinks,citecolor=nicegreen,linkcolor=niceblue}
\hypersetup{colorlinks=true}

\usepackage{pifont}

\allowdisplaybreaks

\setlongtables

\newcommand{\bea}{\begin{eqnarray}}
\newcommand{\eea}{\end{eqnarray}}
\newcommand{\beq}{\begin{equation}}
\newcommand{\eeq}{\end{equation}}
\newcommand{\ec}{\end{center}}
\newcommand{\bc}{\begin{center}}

\newcommand{\pdir}{p\kern -5.2pt\raise 0.2ex\hbox {/}}

\newcommand{\vdir}{v\kern -5.75pt\raise 0.15ex\hbox {/}}
\newcommand{\kdir}{k\kern -5.75pt\raise 0.15ex\hbox {/}}
\newcommand{\epsdir}{\epsilon\kern -5.0pt\raise 0.15ex\hbox {/}}
\newcommand{\bvdir}{\bar{v}\kern -5.75pt\raise 0.15ex\hbox {/}}
\newcommand{\Ddir}{D\kern -7.75pt\raise 0.20ex\hbox {/}}
\newcommand{\Adir}{A\kern -7.75pt\raise 0.20ex\hbox {/}}
\newcommand{\ldir}{l\kern -5.0pt\raise 0.2ex\hbox{/}}
\newcommand{\varepsdir}{\varepsilon\kern -5.5pt\raise 0.15ex\hbox{/}}

\makeatother

\definecolor{niceblue}{rgb}{0.15,0.15,0.6}
\definecolor{nicegreen}{rgb}{0.1,0.5,0.1}
\definecolor{Red}{rgb}{1.,0.,0.}

\definecolor{Green}{rgb}{0.2,.7,0.2}

\begin{document}
\unitlength = 1mm

\thispagestyle{empty} 

\begin{flushright}
\begin{tabular}{l}
{\tt \footnotesize \color{blue}IFT-UAM/CISC-19-56}\\ {\tt \footnotesize \color{blue}FTUAM-19-8}\\
\end{tabular}
\end{flushright}
\begin{center}
\vskip 3.4cm\par
{\par\centering \textbf{\Large   \bf Revisiting the production of ALPs at \boldmath$B$\unboldmath-factories}}
\vskip 1.2cm\par
{\scalebox{.85}{\par\centering \large  
\sc L.~Merlo$^a$, F.~Pobbe$^{b,c}$, S.~Rigolin$^{b,c}$ and O.~Sumensari$^{b}$}
{\par\centering \vskip 0.7 cm\par}
{\sl 
$^a$~{Departamento de F\'isica Te\'orica and Instituto de F\'isica Te\'orica, IFT-UAM/CSIC,
Universidad Aut\'onoma de Madrid, Cantoblanco, 28049, Madrid, Spain}}\\
{\par\centering \vskip 0.25 cm\par}
{\sl 
$^b$~Istituto Nazionale Fisica Nucleare, Sezione di Padova, I-35131 Padova, Italy}\\
{\par\centering \vskip 0.25 cm\par}
{\sl 
$^c$~Dipartamento di Fisica e Astronomia ``G.~Galilei", Universit\`a degli Studi di Padova, Italy}\\

{\vskip 1.65cm\par}}

\end{center}

\vskip 0.85cm
\begin{abstract}
In this paper, the production of Axion-Like Particles (ALPs) at $B$-factories via the process $e^+e^- \to \gamma a$ is revisited. To this purpose, the relevant cross-section is computed via an effective Lagrangian with simultaneous ALP couplings to $b$-quarks and photons. The interplay between resonant and non-resonant contributions is shown to be relevant for experiments operating at $\sqrt{s}=m_{\Upsilon(nS)}$, with $n=1,2,3$, while the non-resonant one dominates at $\Upsilon(4S)$. These effects imply that the experimental searches performed at different quarkonia resonances are sensitive to complementary combinations of ALP couplings. To illustrate these results, constraints from existing BaBar and Belle data on ALPs decaying into invisible final states are derived, and the prospects for the Belle-II experiment are discussed.
\end{abstract}
\newpage
\setcounter{page}{1}
\setcounter{footnote}{0}
\setcounter{equation}{0}
\noindent

\renewcommand{\thefootnote}{\arabic{footnote}}

\setcounter{footnote}{0}


\newpage

\section{Introduction}
\label{sec:intro}

Light pseudoscalar particles naturally arise in many extensions of the Standard Model (SM), including the ones 
endowed with an approximate global symmetry spontaneously broken at a given scale, $f_a$. Sharing a common nature with the QCD axion \cite{Peccei:1977hh,Wilczek:1977pj,Weinberg:1977ma}, (pseudo) Nambu-Goldstone bosons are generically referred to as Axion-Like Particles (ALPs). The ALP mass $m_a$ can, in general, be much lighter than the symmetry breaking scale $f_a$, as it is paradigmatically exemplified in the KSVZ and DFSZ invisible axion models~\cite{Kim:1979if,Shifman:1979if,Zhitnitsky:1980tq,Dine:1981rt}. Therefore, it may be not inconceivable that the first hint of new physics at (or above) the TeV scale could be the discovery of 
a light pseudoscalar state. 

The ALP parameter space has been intensively explored in several terrestrial facilities, covering a wide energy range~\cite{Mimasu:2014nea,Jaeckel:2015jla,Bauer:2017ris,Brivio:2017ije,Alonso-Alvarez:2018irt,Harland-Lang:2019zur,Baldenegro:2018hng}, as well as by many astrophysical and cosmological probes~\cite{Cadamuro:2011fd,Millea:2015qra,DiLuzio:2016sbl}. The synergy of these experimental searches allows to access several orders of magnitude in ALP masses and couplings, cf.~e.g.~Ref.~\cite{Irastorza:2018dyq} and references therein. While astrophysics and cosmology impose severe constraints on ALPs in the sub-KeV mass range, the most efficient probes of weakly-coupled particles in the MeV-GeV range come from experiments acting on the precision frontier~\cite{Essig:2013lka}. Fixed-target facilities such as NA62~\cite{NA62:2017rwk} and the proposed SHiP experiment~\cite{Alekhin:2015byh} can be very efficient to constrain long-lived particles. Furthermore, the rich ongoing research program in the $B$-physics experiments at LHCb~\cite{Aaij:2015tna,Aaij:2016qsm} and the $B$-factories~\cite{Bevan:2014iga,Kou:2018nap} offers several possibilities to probe yet unexplored ALP couplings. 

The main goal of this paper is to re-examine existing BaBar and Belle flavor-conserving constraints on ALPs, and to identify the most promising experimental searches to be performed at the forthcoming Belle-II experiment. While there have been several studies discussing signatures of ALPs at $B$-factories \cite{Masso:1995tw,Dolan:2017osp,CidVidal:2018blh,deNiverville:2018hrc}, many clarifications are still needed. Firstly, the resonant contributions to the ALP production, via the $e^+ e^- \to \Upsilon(nS)\to a\gamma$ process, have been overlooked before. As will be shown, these effects can induce numerically significant corrections to experimental searches performed at $\sqrt{s}=m_{\Upsilon(nS)}$, with $n=1,2,3$. Another improvement provided here concerns the theoretical expression for the $\Upsilon\to \gamma a$ branching fraction. Previous studies estimate this quantity by considering either the ALP coupling to $b$-quarks~\cite{Dolan:2014ska}, or to gauge bosons~\cite{Dolan:2017osp}. In this paper, it will be shown that the simultaneous presence of both interactions, as expected in the most general framework, gives rise to new interesting phenomenological features. 

In order to assess the limits on ALP couplings, one should specify not only the ALP production mechanism, but also its decay products. In this paper, it will be assumed that the ALP does not decay into visible particles. Such a scenario can be easily achieved by assuming a sufficiently large ALP coupling to a stable dark sector, as motivated by several dark matter models. The conclusions related to ALP production are, however, general and they can also be applied to the reinterpretation of experimental searches with visible decays in the detector, as will be discussed in the following.

\

The remainder of this paper is organized as follows. In Sec.~\ref{sec:setup}, the effective Lagrangian describing the interactions between the ALP and SM particles up to dimension five is introduced. In Sec.~\ref{sec:expressions}, the relevant non-resonant and resonant contributions to the \mbox{$e^+ e^- \rightarrow \gamma a$} process are computed. Sec.~\ref{sec:exp} is devoted to the classification of experimental searches that can be performed at $B$-factories. The phenomenological implications of these results are illustrated in Sec.~\ref{sec:constraints}, by reinterpreting present constraints on the benchmark scenario of an ALP decaying into invisible particles, and by discussing the corresponding prospects for the Belle-II experiment. Conclusions and final remarks are presented in Sec.~\ref{sec:summary}.

\section{ALP effective Lagrangian}
\label{sec:setup}

The dimension-five effective Lagrangian describing ALP interactions, above the electroweak symmetry breaking scale, can be generically written as~\cite{Brivio:2017ije}
%
\begin{align}
\begin{split}
\delta\mathcal{L}_{\mathrm{eff}}  =  \dfrac{1}{2} (\partial^\mu a) \,(\partial_\mu a) - \dfrac{m_a^2}{2}a^2 &- 
  \dfrac{c_{aBB}}{4} \dfrac{a}{f_a}\, B^{\mu\nu}\widetilde{B}_{\mu\nu}  
  - \dfrac{c_{aWW}}{4} \dfrac{a}{f_a}\, W^{\mu\nu}\widetilde{W}_{\mu\nu} \\
  & - \dfrac{c_{agg}}{4} \dfrac{a}{f_a}\, G_{a}^{\mu\nu}
      \widetilde{G}^a_{\mu\nu} - \dfrac{\partial_\mu a}{2 f_a} \sum_f c_{aff}\, \overline{f}\gamma^\mu \gamma_5 f \,,
  \label{eq:leff-alp-0}
\end{split}
\end{align}
%
where $\widetilde{V}^{\mu\nu}=\frac{1}{2} \,\varepsilon^{\mu\nu\alpha\beta} V_{\alpha \beta}$, $c_{aff}$ and $c_{aVV}$ denote the ALP couplings to fermions and to the SM gauge bosons,~$V\in \lbrace g,B,W \rbrace$, respectively. The ALP mass $m_a$ and the scale $f_a$ are assumed to be independent parameters, in contrast to the QCD-axion paradigm, which is characterized by the relation $m_a \, f_a \approx m_\pi \, f_\pi$~\cite{Kim:2008hd}. Moreover, if the ultraviolet completion of Eq.~\eqref{eq:leff-alp-0} is not specified, the ALP couplings to gauge bosons and fermions are described by independent parameters, which can be of the same order of magnitude and which should, therefore, be simultaneously considered in phenomenological analyses.

At the energy-scales relevant at $B$-factories, the ALP interactions with the $Z$ boson can be safely neglected, due to the Fermi constant suppression. Furthermore, the  ALP couplings to the top-quark and $W^\pm$ boson are relevant only to the study of flavor-changing neutral currents observables, which are complementary to the probes discussed here -- see e.g.~Refs.~\cite{Freytsis:2009ct,Izaguirre:2016dfi,Dobrich:2018jyi,Gavela:2019wzg} for a recent discussion. The only relevant couplings in Eq.~\eqref{eq:leff-alp-0} at low-energies are
\begin{align}
\begin{split}
\delta\mathcal{L}_{\mathrm{eff}} \, \supset \, &\dfrac{1}{2} (\partial^\mu a) \, (\partial_\mu a) - \dfrac{m_a^2}{2}a^2 \\[0.25em]
&-    \dfrac{c_{a\gamma\gamma}}{4 } \dfrac{a}{f_a} F_{\mu\nu}\widetilde{F}_{\mu\nu} 
  - \dfrac{c_{agg}}{4} \dfrac{a}{f_a}\, G_{a}^{\mu\nu} \widetilde{G}^a_{\mu\nu} - \dfrac{\partial_\mu a}{2 f_a} \sum_f c_{aff}\, \overline{f}\gamma^\mu \gamma_5 f\,,
\end{split}
\label{eq:leff-alp}
\end{align}
where $c_{a\gamma\gamma} = c_{aBB}\, \cos^2 \theta_W + c_{aWW}\, \sin^2 \theta_W$. The couplings relevant to ALP production are $\lbrace c_{a\gamma\gamma}, c_{abb} \rbrace$, while the other couplings only contribute to the ALP branching fractions. 

Light pseudoscalar particles can also act as portals to a light dark sector~\cite{Boehm:2014hva,Arcadi:2018pfo}. In this case, to describe these additional interactions, new couplings are customarily introduced. By assuming, for instance, an extra light and neutral dark fermion state $\chi$, the following term should be considered in the effective Lagrangian:
\bea
 \delta\mathcal{L}_{\mathrm{eff}} \, \supset \, - c_{a\chi\chi}\,\dfrac{\partial_\mu a}{2 f_a} \,\overline{\chi}\gamma^\mu \gamma_5 \chi\,,
\eea
where $c_{a\chi\chi}$ denotes a generic coupling, which can induce a sizable ALP decay into invisible final states, as will be considered in the following. 

In the remained of this paper, $c_{aii} \equiv 
c_{aii}^{\mathrm{eff}}(\mu=m_b)$ will be assumed, and the ALP mass will be taken in the range $m_a\in(0.1-10)$ GeV, for which $B$-factories provide some of the most stringent bounds on its couplings.

\boldmath
\section{$B$-factories probes of invisible ALPs}
\label{sec:expressions}
\unboldmath

In this Section, the potential of $B$-factories to probe ALP couplings in the $e^+ e^- \to \gamma a$ channel will be discussed. 
Two main scenarios are typically considered in the literature, depending on the relative strength of the ALP coupling to SM and dark sector particles: either $|c_{a\chi\chi}| \ll |c_{a\mathrm{SM}}|$, or $|c_{a\chi\chi}| \gg |c_{a \mathrm{SM}}|$. In the first case, for $m_a$ values in the GeV range, the ALP would typically decay in the detector, leaving the signatures $\gamma a (\to jj)$ \cite{Lees:2013vuj,
Lees:2015jwa}, $\gamma a(\to\gamma\gamma)$ \cite{delAmoSanchez:2010ac,Seong:2018gut,Aubert:2008as} and $\gamma a (\to \ell\ell)$ \cite{Aubert:2009cp,Lees:2012iw,Lees:2012te}, with $\ell= \lbrace e,\mu,\tau\rbrace$. This scenario is dubbed the \textit{visible} ALP. If, 
however, the coupling to the dark sector $c_{a\chi\chi}$ is large, in comparison to the SM couplings, then the ALP will decay 
predominantly into an invisible channel, providing the mono-$\gamma$ plus missing energy signature. This scenario will be referred to as the \textit{invisible} ALP,~\footnote{The \textit{invisible} ALP case should not be confused with the traditional \textit{invisible} QCD axion~\cite{Kim:1979if,Shifman:1979if,Zhitnitsky:1980tq,Dine:1981rt}.} which also covers the possibility of a sufficiently long-lived ALP that does not decay in the detector.

In this paper, the \textit{invisible ALP} scenario will be considered for the sake of illustration. The main goal will be (i) to revisit the theoretical expressions available in the literature, including ALP coupling to bottom quarks, as well as 
previously unaccounted experimental uncertainties, and (ii) to propose an optimal strategy for future ALP analyses. Even though the main focus will be the minimalistic \textit{invisible} ALP scenario, most of the observations that will be made in this paper can be translated \textit{mutatis mutandis} to the \textit{visible} case.

\begin{figure}[t]
\centering
\includegraphics[width=.4\linewidth]{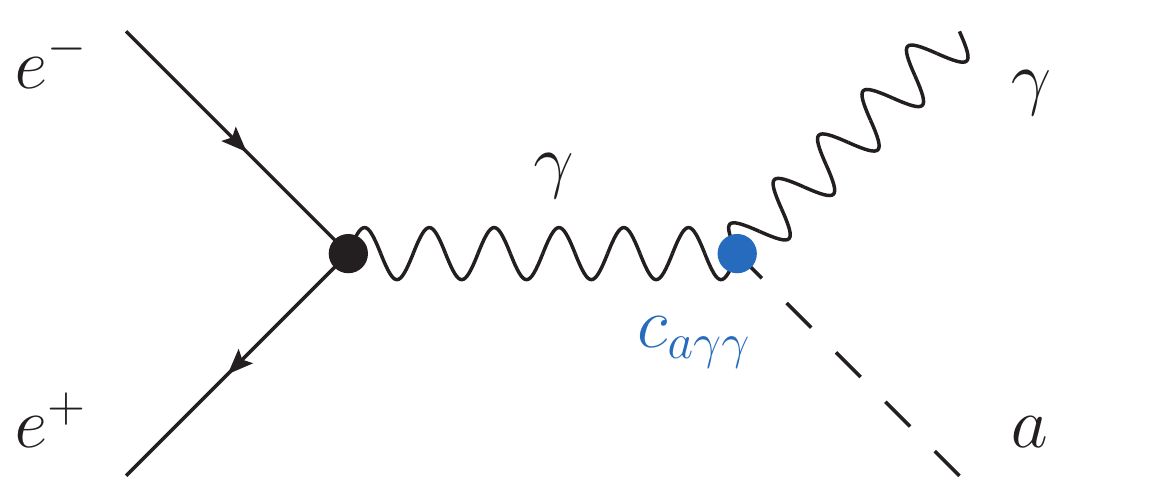}
\caption{\small \sl Non-resonant contribution to the process $e^+ e^-\to \gamma a$ produced via the effective coupling 
$c_{a\gamma\gamma}$ defined in Eq.~\eqref{eq:leff-alp}.}
\label{fig:diag-non-resonant}
\end{figure}

\paragraph*{\bf Non-resonant ALP Production.} 
The most straightforward way of producing ALPs in $e^+e^-$ facilities is via the non-resonant process $e^+ e^- \to \gamma a$, 
as illustrated in Fig.~\ref{fig:diag-non-resonant}. 
If the ALP does not decay inside the detector, as assumed throughout the paper, this process would result in an energetic $\gamma$ 
plus missing energy. The differential cross-section for this process, keeping explicit the ALP mass dependence, can be expressed 
as~\cite{Marciano:2016yhf}
\begin{equation}
\label{eq:dsigma-monogamma}
\left(\dfrac{\mathrm{d}\sigma(s)}{\mathrm{d}\cos\theta_\gamma} \right)_{\text{NR}} = 
  \dfrac{\alpha_{\mathrm{em}}}{128} \dfrac{c_{a\gamma\gamma}^2}{f_a^2} \left(3+ \cos 2\theta_\gamma\right)
  \left(1- \dfrac{m_a^2}{s}\right)^3\,,
\end{equation}
where $s=E_{\mathrm{cm}}^2$, and $\theta_\gamma$ is the angle of photon emission with respect to the collision axis, in the 
center-of-mass frame. In this expression, the contributions coming from the exchange of an off-shell $Z$ boson, which are also 
induced by $c_{aWW}$ in Eq.~\eqref{eq:leff-alp-0}, have been neglected, since they are suppressed, at low-energies, by $s/m_Z^2 \ll 1$. 
The integrated cross-section then gives:
\begin{equation}
\label{eq:sigma-monogamma}
\sigma_{\text{NR}} (s)= \dfrac{\alpha_{\mathrm{em}}}{24} \dfrac{c_{a\gamma\gamma}^2}{f_a^2}\left(1- \dfrac{m_a^2}{s}\right)^3\,.
\end{equation}
While the non-resonant contribution to ALP production given above is unavoidable in any experiment relying on $e^+ e^-$ collisions 
\cite{Masso:1995tw}, the situation at $B$-factories is more intricate since these experiments operate at specific $\Upsilon(nS)$ 
resonances. Therefore, it is crucial to account for the resonantly enhanced contributions, which can be numerically 
significant, as will be shown in the following. 

%
\paragraph*{\bf Resonant ALP Production.}

\begin{figure*}[!h]
\centering
\includegraphics[scale=.88]{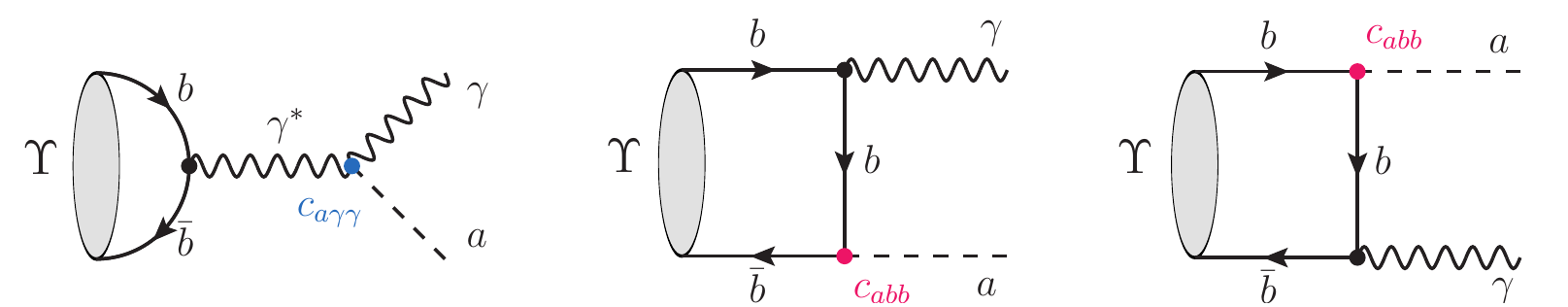}
\caption{\small \sl Contributions to the $\Upsilon(nS)\to \gamma a$ decays from the effective couplings introduced in the 
Lagrangian or Eq.~(\eqref{eq:leff-alp}).}
\label{fig:diag-resonant}
\end{figure*}

Vector quarkonia can produce significant resonant contributions to the mono-$\gamma$ channel, $e^+e^- \to \Upsilon \to \gamma a$, since they are very narrow particles coupled to the electromagnetic current. Assuming a fixed center-of-mass energy $\sqrt{s} \approx m_\Upsilon$, as is the case at $B$-factories, and using the Breit-Wigner approximation, one finds for the resonant cross-section 
\begin{equation}
\label{eq:res-xsection}
\sigma_\mathrm{R}(s) = \sigma_{\mathrm{peak}}\,\dfrac{ m_\Upsilon^2\Gamma_\Upsilon^2}{(s-m_\Upsilon^2)^2+m_\Upsilon^2 
\Gamma_\Upsilon^2}\,\mathcal{B}(\Upsilon\to \gamma a)\,, 
\end{equation}
where $m_\Upsilon$ and $\Gamma_\Upsilon$ are the mass and width of a specific $\Upsilon$ resonance, and $\sigma_{\mathrm{peak}}$ 
is the peak cross-section defined as
\begin{equation}
\sigma_{\mathrm{peak}} = \dfrac{12 \pi \mathcal{B}(\Upsilon \to ee)}{m_\Upsilon^2}\,,
\end{equation}
with $\mathcal{B}(\Upsilon \to ee)$ being the leptonic branching fraction, experimentally determined for the different $\Upsilon(nS)$ 
resonances~\cite{Tanabashi:2018oca}. The effective couplings defined in Eq.~\eqref{eq:leff-alp} appear, instead, in 
the $\mathcal{B}(\Upsilon\to \gamma a)$ branching fraction, as illustrated in Fig.~\ref{fig:diag-resonant}, which will be computed in full generality in Secs.~\ref{sec:fomulas-1} and \ref{sec:fomulas-2}.

%
\paragraph*{\bf Non-resonant vs Resonant ALP Production.}

Naively, one would expect that the resonant cross-section {(Eq.~\eqref{eq:res-xsection})} clearly dominates over the non-resonant one {(Eq.~\eqref{eq:sigma-monogamma}) for the very narrow $\Upsilon(1S)$, $\Upsilon(2S)$ and $\Upsilon(3S)$ resonances, since $\Gamma_\Upsilon/m_\Upsilon \ll 1$. Nevertheless, this turns out to not be the case at $B$-factories, since these experiments are 
intrinsically limited by the energy spread of the $e^+e^-$ beam, which is of order $\sigma_W \approx 5$~MeV at current facilities.~\footnote{More specifically, the energy spread was $\sigma_W = 5.5$~MeV at BaBar (PEP) \cite{Tanabashi:2018oca} and $\sigma_W=5.24$~MeV at Belle (KEKB)~\cite{Koiso:2013tgf}, and it is expected to be $\sigma_W = 5.45$~MeV at Belle-II (SuperKEKB)~\cite{Ohnishi:2013fma}.} This value is considerably larger than the width of 
these resonances, which therefore cannot be fully resolved at $B$-factories. The only exception 
is the $\Upsilon(4S)$ resonance, for which $\Gamma_{\Upsilon(4S)}=20.5$~MeV~\cite{Aubert:2004pwa}. Therefore, one should expect a sizable reduction of the estimation in Eq.~\eqref{eq:res-xsection} for the lightest quarkonia resonances, due to this intrinsic experimental uncertainty. 

To account for the beam-energy uncertainties in Eq.~\eqref{eq:res-xsection}, the procedure presented in Ref.~\cite{Eidelman:2016aih} has been adopted by performing a convolution of $\sigma_{\mathrm{R}}(s)$ with a Gaussian distribution, with spread $\sigma_W$,
\begin{equation}
\label{eq:prod-vis}
\langle \sigma_\mathrm{R} (s) \rangle_\mathrm{vis} = \int \! \! \mathrm{d}q \, \dfrac{\sigma_\mathrm{R} (q^2)}{\sqrt{2\pi}\sigma_W} 
\mathrm{exp}\left[-\dfrac{(q-\sqrt{s})^2}{2\sigma_W^2}\right]\,.
\end{equation}
At the very narrow $\Upsilon(nS)$ resonances, with $n=1,2,3$, one finds $\Gamma_\Upsilon\ll \sigma_W$, in such a way that the previous 
expression can be simplified by writing~\cite{Eidelman:2016aih}
\begin{equation}
\label{eq:prod-ups}
\langle \sigma_\mathrm{R} (m_\Upsilon^2) \rangle_\mathrm{vis} = \rho \,\sigma_{\mathrm{peak}}\,\mathcal{B}(\Upsilon\to \gamma a)\,,
\end{equation}
where the parameter $\rho$, defined as
\begin{equation}
\rho=\sqrt{\dfrac{\pi}{8}}\dfrac{\Gamma_\Upsilon}{\sigma_W}\,,
\end{equation}
accounts for the cross-section suppression at the peak due to the finite beam-energy spread. These effects will be quantified in the following in two scenarios: (i) ALP with predominant couplings to photons, $|c_{a\gamma\gamma}|\gg |c_{abb}|$, and (ii) the general case with both $c_{a\gamma\gamma}$ and $c_{abb}$ nonzero.

\subsection{The photo-philic scenario: $|c_{a\gamma\gamma}| \gg |c_{abb}|$}
\label{sec:fomulas-1}

The scenario most commonly considered in the literature is the one with predominant ALP couplings to photons~\cite{Dolan:2017osp}. In this case, by neglecting $c_{abb}$, the first diagram in Fig.~\eqref{fig:diag-resonant} leads to 
\begin{equation}
\label{eq:ups-gamma}
\mathcal{B}(\Upsilon\to \gamma a)\Big{\vert}_{c_{abb}=0} =\dfrac{m_\Upsilon^2 }{32 \pi \alpha_{\mathrm{em}}}
\dfrac{c_{a\gamma\gamma}^2}{f_a^2} \left(1-\dfrac{m_a^2}{m_\Upsilon^2} \right)^3 \mathcal{B}(\Upsilon\to ee)\,,
\end{equation}
which agrees with Ref.~\cite{Masso:1995tw} in the massless ALP limit. Note that this expression does not require assumptions on hadronic uncertainties, since the hadronic matrix element appearing in this computation, namely $\langle 0 | \bar{b}\gamma^\mu b| \Upsilon (p)\rangle$, also enters the process $\Upsilon\to ee$ which has been accurately measured experimentally~\cite{Tanabashi:2018oca}. Alternatively, $\mathcal{B}(\Upsilon\to ee)$ can be expressed in terms of the $\Upsilon$ decay constant, defined as
\begin{equation}
\label{eq:ups-decayconstant}
\langle 0 | \bar{b}\gamma^\mu b| \Upsilon (p)\rangle \equiv m_\Upsilon\, f_{\Upsilon}\, \varepsilon^\mu (p)\,,
\end{equation}
which encapsulates the QCD dynamics of this process and which has been independently computed, for the lighter $\Upsilon$ resonances, by means of numerical simulations of QCD on the lattice~\cite{Gray:2005ur,Lewis:2011ti,Abada:2015zea}. The previous definition allows to recast Eq.~\eqref{eq:ups-gamma} in the more convenient form,
\begin{equation}
\mathcal{B}(\Upsilon \to \gamma a)\Big{\vert}_{c_{abb}=0} = \dfrac{\alpha_{\mathrm{em}}}{216\, \Gamma_\Upsilon} 
m_\Upsilon f_\Upsilon^2 \dfrac{c_{a\gamma\gamma}^2}{f_a^2} \left(1-\dfrac{m_a^2}{m_{\Upsilon}^2}\right)^3\,.
\end{equation}
The Lattice QCD (LQCD) determinations of $f_{\Upsilon(nS)}$ are summarized in Tab.~\ref{tab:fy} along with the values extracted from the experimentally determined $\mathcal{B}(\Upsilon(nS)\to e^+e^-)$~\cite{Tanabashi:2018oca}, showing a reasonable agreement.

\

\begin{table*}[!h]
  \centering
  \renewcommand{\arraystretch}{1.3}
\begin{tabular}{|c|c|c|}
\hline
\hspace{0.25cm} $\Upsilon(nS)$ \hspace{0.25cm} & \hspace{0.25cm} $f_{\Upsilon}^{\mathrm{latt.}}~(\mathrm{MeV})$ \hspace{0.25cm} & 
\hspace{0.25cm} $f_{\Upsilon}^{\mathrm{exp.}}~(\mathrm{MeV})$ \hspace{0.25cm} \\
\hline\hline
$\Upsilon(1S)$  & $680(14)$ & $659(17)$ \\
$\Upsilon(2S)$  & $494(15)$ & $468(27)$ \\
$\Upsilon(3S)$  & $539(84)$ & $405(26)$ \\ 
$\Upsilon(4S)$  & -- & $349(23)$ \\
\hline
\end{tabular}
\caption{ \em $\Upsilon(nS)$ decay constants computed by means of numerical lattice simulations~\cite{Gray:2005ur,
Lewis:2011ti,Abada:2015zea} or determined experimentally from $\mathcal{B}(\Upsilon(nS)\to e^+e^-)$~\cite{Tanabashi:2018oca}.}  
\label{tab:fy}
\end{table*}

\begin{table*}[!t]
\centering
  \renewcommand{\arraystretch}{1.4} 
\begin{tabular}{|c|cc|cc||c|}
\hline
$\Upsilon(nS)$  & $m_\Upsilon~[\mathrm{GeV}]$ &  $\Gamma_\Upsilon~[\mathrm{keV}]$  & 
$\sigma_{\mathrm{peak}}$~[nb] & $\rho$  & $\langle\sigma_{\mathrm{R}}(m_\Upsilon^2)\rangle_{\mathrm{vis}}/\sigma_{\text{NR}}$\\
\hline\hline
$\Upsilon(1S)$  & $9.460$  & $54.02$  & $3.9(18)\times 10^{3}$             & $6.1\times 10^{-3}$ &  $0.53(5)$ \\
$\Upsilon(2S)$  & $10.023$ &  $31.98$  & $2.8(2)\times 10^{3}$             & $3.7\times 10^{-3}$ &  $0.21(3)$ \\
$\Upsilon(3S)$  & $10.355$ & $20.32$  & $3.0(3)\times 10^{3}$              & $2.3\times 10^{-3}$ &  $0.16(3)$ \\ 
$\Upsilon(4S)$  & $10.580$ & $20.5\times 10^3$  &  $2.10(10)$ & $0.83$    & $3.0(3)\times 10^{-5}$ \\
\hline
\end{tabular}
\caption{\em Estimated \textit{visible} cross-section at Belle-II for $e^+e^-\to \Upsilon \to \gamma a$ compared to the 
non-resonant one, $e^+e^-\to \gamma^\ast \to \gamma a$. Here, vanishing ALP couplings with $b$-quarks have been assumed, $c_{abb} =0$. 
Experimental inputs are taken from Ref.~\cite{Tanabashi:2018oca}. Belle-II machine parameter have been considered~\cite{Ohnishi:2013fma}, namely  $\sigma_W = 5.45$~MeV for the beam-energy spread. }  
\label{tab:xsection}
\end{table*}

In Table~\ref{tab:xsection}, Eq.~\eqref{eq:ups-gamma} is combined with Eqs.~\eqref{eq:sigma-monogamma} and \eqref{eq:prod-ups}} to estimate the resonant and non-resonant cross-section, for each $\Upsilon$ resonance, along with the peak cross-section $\sigma_{\mathrm{peak}}$ and the suppression parameter $\rho$. This computation has been performed with the Belle-II (KEKB) energy-spread for illustration, which is similar to the ones from BaBar (PEP) and Belle (KEK). From this table, one learns that even though the peak 
cross-section is large for the $\Upsilon(nS)$ resonances ($n=1,2,3$), the beam-energy uncertainties entail a considerable suppression of 
the \textit{visible} cross-section. These effects are milder for the $\Upsilon(4S)$ resonance, but in turn the cross-section at the peak is much smaller in this case. The final results are summarized in the last column of Table~\ref{tab:xsection}, which shows that the effective resonant cross-section is smaller than the non-resonant one, but it still contributes with numerically significant effects. For the (very) narrow resonances $\Upsilon(nS)$ ($n=1,2,3$), the resonant contribution amounts to corrections between $20\%$ and $50\%$ 
to the non-resonant one, which should be included when reinterpreting experimental searches.~\footnote{Interference effects between the 
non-resonant and resonant $c_{a\gamma\gamma}$ terms turn out to be negligible due to the small width of the $\Upsilon(nS)$ resonances.}  On the other hand, for the $\Upsilon(4S)$ resonance the resonant contribution turns out to be negligible, due to its larger width, as expected.

\subsection{The general case: $c_{a\gamma\gamma}\neq 0$ and $c_{abb}\neq 0$}
\label{sec:fomulas-2}

The previous discussion implies that the resonant contributions are not only important to correctly assess limits on the ALP coupling to photons, $c_{a\gamma\gamma}$, but they also open the window to probe the ALP coupling to $b$-quarks, $c_{abb}$, cf.~Fig.~\ref{fig:diag-resonant}. The simultaneous presence of these contributions gives rise to a rich phenomenology which will be discussed in the following.

Firstly, the hadronic matrix element needed to estimate the $c_{abb}$ contribution to $\mathcal{B}(\Upsilon \to \gamma a)$ is 
far more intricate than the one given in Eq.~\eqref{eq:ups-decayconstant}, since this is a QCD-structure dependent emission, 
as depicted in the last two diagrams in Fig.~\eqref{fig:diag-resonant}. This contribution was first computed 
by Wilczek for a SM-like Higgs by using a non-relativistic approximation \cite{Wilczek:1977pj,Wilczek:1977zn}, see 
also Ref.~\cite{Ellis:1979jy,Pantaleone:1984ug,Haber:1978jt,Fayet:2008cn}.~\footnote{Compatible results have also been obtained in Ref.~\cite{Dulian:1987xd} for small pseudoscalar masses by using a 
QCD sum-rules approach.} By using a similar approach, the total $\mathcal{B}(\Upsilon \to \gamma a)$ branching fraction reads
\begin{align}
\label{eq:ups-general}
\begin{split}
\mathcal{B}(\Upsilon \to \gamma a) = &\dfrac{\alpha_\mathrm{em}}{216\,\Gamma_\Upsilon} {m_\Upsilon f_\Upsilon^2 }
\left(1-\dfrac{m_a^2}{m_\Upsilon^2}\right) \, \Bigg{[} \dfrac{c_{a\gamma\gamma}}{f_a} 
\left(1-\dfrac{m_a^2}{m_\Upsilon^2}\right)- 2\dfrac{c_{abb}}{f_a} \Bigg{]}^2\,.
\end{split}
\end{align}
This expression includes, for the first time, the most general $c_{a\gamma\gamma}$ and $c_{abb}$ contributions, as well as their interference. Note, however, that the computation of the $c_{abb}$ contributions are done within a first approximation that considerably simplifies the QCD structure-dependent emission of this decay. If a new physics signal is indeed observed in such observable, a more accurate theoretical calculation would be needed to fully assess the (non-perturbative) effects associated to the last two diagrams in Fig.~\ref{fig:diag-resonant}.

As shown in Eq.~\eqref{eq:ups-general}, the $c_{a\gamma\gamma}$ and $c_{abb}$ couplings can induce comparable contributions to the non-resonant cross-section in Eq.~\eqref{eq:res-xsection}. Moreover, depending on the relative sign of these two couplings, these couplings can interfere destructively or constructively, as will be illustrated with a concrete example in Sec.~\ref{sec:constraints}. Finally, note that Eq.~\eqref{eq:ups-general} shows a different dependence on $m_a$ and $\lbrace c_{a\gamma\gamma}, c_{abb} \rbrace$ than the non-resonant cross-section in Eq.~\eqref{eq:sigma-monogamma}. A comparison between $\langle \sigma_\mathrm{R}\rangle_{\mathrm{vis}}$ and $\langle \sigma_{\mathrm{NR}}\rangle_\mathrm{vis} \approx \sigma_{\mathrm{NR}}$ is postponed to Sec.~\ref{sec:constraints} where a concrete scenario will be considered.

\section{Summary of experimental searches}
\label{sec:exp}

From the previous discussion, one learns that the non-resonant cross-section, via the coupling $c_{a\gamma\gamma}$, is the largest one, but it can be of the same order of the resonant one, cf.~Tab.~\ref{tab:xsection}. Moreover, the latter searches have the advantage of being sensitive to both $c_{a\gamma\gamma}$ and $c_{abb}$ couplings. Based on these observations, ALP searches at $B$-factories can be classified in the following three categories:

\begin{itemize}[leftmargin=*]
\item[\bf i)] {\bf Resonant searches}: Excited quarkonia states $\Upsilon (nS)$ (with $n>1$) can decay into lighter $\Upsilon (nS)$ resonances via pion emission, as for example $\Upsilon (2S) \to \Upsilon(1S)\,\pi^+ \pi^-$ and $\Upsilon (3S) \to \Upsilon(1S)\,\pi^+ \pi^-$. By exploiting the kinematics of these processes one can reconstruct the $\Upsilon(1S)$ meson and then study its decay into a specific final state, which can, for instance, be the invisible $\Upsilon$ decay~\cite{Aubert:2009ae}, or the $\Upsilon$ decay into photon and a light (pseudo)scalar particle \cite{delAmoSanchez:2010ac,Seong:2018gut}. These searches are dubbed \textit{resonant}, since they allow to directly probe $\mathcal{B}(\Upsilon\to \gamma a)$ in a \textit{model-independent} way, regardless of the non-resonant contribution from Fig.~\ref{fig:diag-non-resonant}. In other words, reported limits on $\mathcal{B}(\Upsilon(1S)\to\gamma a)$ can be used to constrain both $c_{a\gamma\gamma}$ and $c_{abb}$ via Eq.~\eqref{eq:ups-general}. Searches along these lines have been performed, for instance, by BaBar 
\cite{delAmoSanchez:2010ac} and, more recently, by Belle~\cite{Seong:2018gut}, under the assumption that the ALP does not decay into visible particles inside the detector.

\item[\bf ii)] {\bf Mixed (non-)resonant searches}: Alternatively, experimental searches could be performed at $\Upsilon(nS)$ (with $n=1,2,3$) without identifying the $\Upsilon$ decay from a secondary vertex. Example of such experimental searches are the ones performed at $\sqrt{s}= m_{\Upsilon(3S)}$~\cite{Aubert:2008as}, where limits on $\mathcal{B}(\Upsilon(3S)\to \gamma a)\times \mathcal{B}(a\to \mathrm{inv})$ are extracted from the total $e^+e^- \to \gamma a (\to \mathrm{inv})$ cross-section. From the above discussion, however, it is clear that this method is probing both resonant (Eq.~\eqref{eq:prod-ups}) and non-resonant (Eq.~\eqref{eq:sigma-monogamma}) cross-sections and therefore model-independent limits on $\mathcal{B}(\Upsilon (3S)\to \gamma a)$ could not be extracted from these experimental results. The only scenarios for which such limits can be derived are the ones with $|c_{a\gamma\gamma}| \ll |c_{abb}|$, as predicted in models with an extended Higgs sector~\cite{McElrath:2005bp,Dermisek:2006py,Domingo:2008rr}, since the non-resonant cross-section vanishes in this case. 

In the most general ALP scenario, instead, the limits on $\lbrace c_{a\gamma\gamma}, c_{abb} \rbrace$ can be obtained from Ref.~\cite{Aubert:2008as} via a rescaling factor,
\begin{align}
\label{eq:xsection-total}
\begin{split}
\dfrac{\langle\sigma_{\text{R}}(s) + \sigma_{\text{NR}}(s)  \rangle_{\mathrm{vis}}}{\langle\sigma_{\mathrm{R}} \rangle_{\mathrm{vis}}} 
&\approx 1+\dfrac{\sigma_{\text{NR}}}{\langle\sigma_{\mathrm{R}}\rangle_\mathrm{vis}} \,,
\end{split}
\end{align}
\noindent which accounts for the non-resonant contributions (Eq.~\eqref{eq:prod-vis}) that have been overlooked experimentally in the total cross-section. For instance, in the case where $c_{abb}= 0$, one obtains constraints on $c_{a\gamma\gamma}$ which are a factor of $\approx 3 $ more stringent than the estimation that overlooks the latter effects, cf.~Table~\ref{tab:xsection}. Note, also, that similar effects have also been overlooked in reinterpretations of other experimental limits, as for example the ones on $\mathcal{B}(\Upsilon(3S)\to \gamma a)\times \mathcal{B}(a\to \mathrm{hadrons})$~\cite{Lees:2013vuj} to constrain the product of ALP couplings to photons and gluons \cite{CidVidal:2018blh}. 

The reinterpretation described above, for the results from Ref.~\cite{Aubert:2008as} and similar searches, has a possible caveat related to the treatment of the background. In these experimental analyses, the background is determined by considering an independent data sample collected outside the resonance region, typically $\approx 30$~MeV below $m_{\Upsilon(3S)}$. While this strategy allows for a robust determination of the SM background in scenarios with $c_{a\gamma\gamma}=0$, this is not an efficient method if $c_{a\gamma\gamma}$ is non-negligible. In the latter case, the background sample also receives contributions from the non-resonant diagram in Fig.~\ref{fig:diag-non-resonant}, which turns out to be the dominant effect. For that reason, it is important that future experimental searches determine the background without relying on off-resonance samples, as performed, for instance, in dark photon searches~\cite{Lees:2017lec}. Furthermore, it would be helpful to also report the limits on the $e^+ e^-\to \gamma a$ cross-section instead of the $\Upsilon(nS)$ branching fraction, as these results contain the full information on both resonant and non-resonant contributions.

\item[\bf iii)] {\bf Non-resonant searches}: The resonant cross-section is negligible at the $\Upsilon(4S)$ resonance, as can be 
seen from Table~\ref{tab:xsection}, since its mass lies just above the $B \overline{B}$ production threshold. Therefore, experimental searches at the $\Upsilon(4S)$ resonance can only probe the $c_{a\gamma\gamma}$ coupling via the 
non-resonant ALP production illustrated in Fig.~\ref{fig:diag-non-resonant}. To our knowledge, no such ALP search has been performed yet at 
$B$-factories. For the future, this type of search could exploit the large luminosity collected at $\Upsilon(4S)$ Belle-II, providing 
the most stringent limits on $c_{a\gamma\gamma}$ for a GeV ALP mass. See Ref.~\cite{Dolan:2017osp} for a recent discussion on Belle-II prospects. 

\end{itemize} 

In summary, ALP production receives both resonant and non-resonant contributions at $B$-factories. The interplay between these production mechanisms 
allows to classify three complementary experimental strategies: (i) \textit{resonant} searches of $\Upsilon \to \gamma a$, from which one could infer bounds on $c_{abb}$ and $c_{a\gamma\gamma}$, (ii) \textit{mixed (non-)resonant} searches which are sensitive to a different combination of $c_{abb}$ and $c_{a\gamma\gamma}$, and (iii) \textit{non-resonant} searches which depend solely on $c_{a\gamma\gamma}$. Before 
deriving constraints on the ALP couplings from existing BaBar and Belle data, it is important to stress once again that the conclusions outlined above are general and that they apply, for instance, to searches for ALP decaying into visible particles, such as hadrons~\cite{Lees:2013vuj,
Lees:2015jwa}, $\mu\mu$~\cite{Aubert:2009cp,Lees:2012iw} and $\tau\tau$ \cite{Lees:2012te}.

\begin{figure*}[t!]
\centering
\includegraphics[width=.5\linewidth]{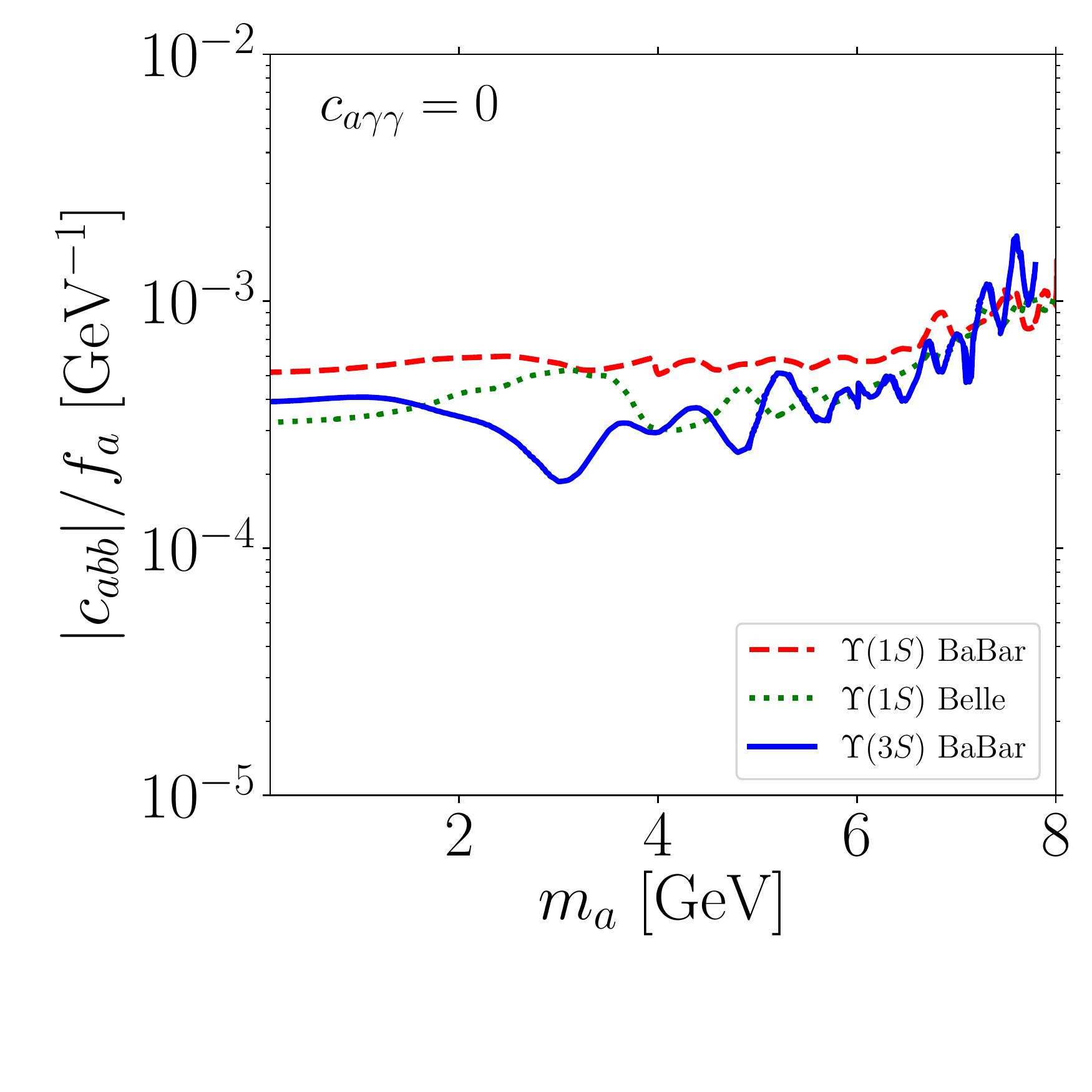}~\includegraphics[width=.5\linewidth]{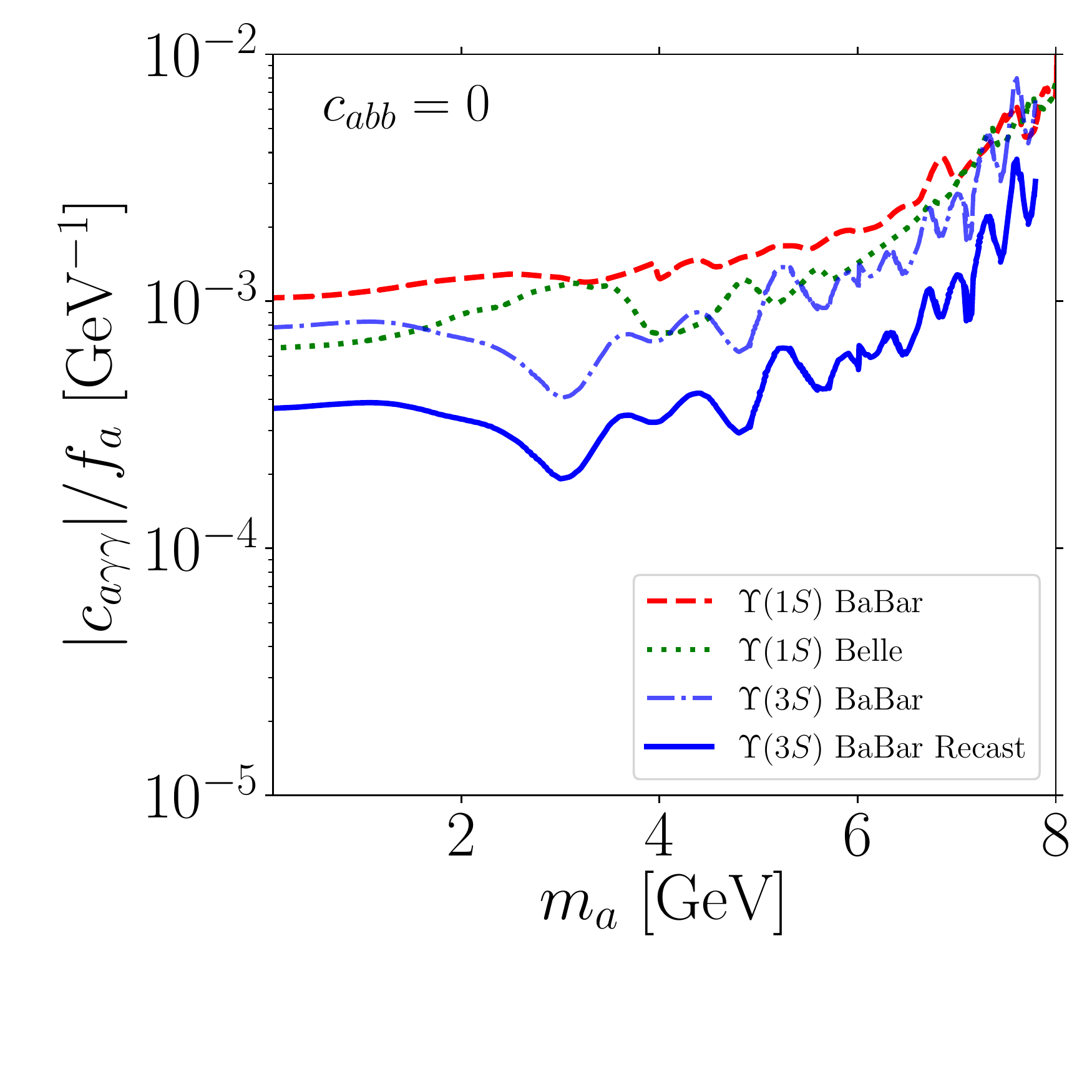} 
\caption{\em \small \sl Upper limits on $c_{abb}/f_a$ (left panel) and $c_{a\gamma\gamma}/f_a$ (right panel) as a function of $m_a$ for the invisible ALP scenario. Experimental limits on $\mathcal{B}(\Upsilon (nS)\to \gamma A)\times\mathcal{B}(a\to \mathrm{inv})$ obtained by BaBar~\cite{delAmoSanchez:2010ac,Aubert:2008as} and Belle~\cite{Seong:2018gut} are considered. For the constraint on $c_{a\gamma\gamma}$ obtained from data 
collected at $\Upsilon(3S)$ resonance~\cite{Aubert:2008as}, the reinterpretation of BaBar limits that neglects the non-resonant ALP production (blue dashed-dotted line) is also considered, along with the rescaled limit that accounts for both resonant and non-resonant ALP production (solid blue line), cf.~Eq.~\eqref{eq:xsection-total}. The latter results provide the most stringent limits on the ALP photon coupling from searches at $B$-factories.}
\label{fig:Babar-bounds}
\end{figure*}

\section{Constraining the ALP parameter space}
\label{sec:constraints}

In this Section, constraints on the ALP parameter space are derived from existing BaBar and Belle data, and prospects for the Belle-II experiment are discussed. For illustration, the \emph{invisible} ALP scenario will be considered, by assuming that $\mathcal{B}(a\to \mathrm{inv})= 1$, or equivalently, that the ALP does not decay inside the detector. As anticipated above, the results derived below can be easily recast to scenarios in which the invisible ALP branching fraction is smaller than one.

Firstly, separate constraints on $c_{a\gamma\gamma}$ and $c_{abb}$ are derived by assuming that the other Wilson coefficient vanishes. These couplings are subject to the limits on $\mathcal{B}(\Upsilon(1S) \to \gamma a)\times \mathcal{B}(a\to \mathrm{inv})$ reported by BaBar~\cite{delAmoSanchez:2010ac} and Belle~\cite{Seong:2018gut}, in which the quarkonia state is produced via the $\Upsilon(2S)\to \Upsilon(1S)\pi^+\pi^-$ decay, cf.~discussion in Sec.~\ref{sec:exp}. Limits on $\mathcal{B}(\Upsilon(3S) \to \gamma a)\times \mathcal{B}(a\to \mathrm{inv})$ reported by BaBar~\cite{Aubert:2008as} are also considered by including the non-resonant contribution overlooked in the experimental analysis, cf.~Eq.~\eqref{eq:xsection-total}. These constraints are combined in Fig.~\ref{fig:Babar-bounds} to constrain $c_{a\gamma\gamma}/f_a$ and $c_{abb}/f_a$ as a function of the ALP mass. While the limits on $c_{abb}$ from the different experimental searches turn out to be similar, the recast described above of $\Upsilon(3S)$ data provides the most stringent limit on $c_{a\gamma\gamma}$. For comparison, the limits obtained by neglecting the non-resonant contribution are also depicted in the same plot by the dashed-dotted line, which turn out to be weaker, as expected. It should be stressed that this reinterpretation is not strictly correct due to the background treatment in Ref.~\cite{Aubert:2008as}, but it can be seen as the expected sensitivity of such searches if the background is determined without relying on off-resonance samples, as discussed in Sec.~\ref{sec:exp}.

\begin{figure*}[t!]
\centering
\includegraphics[width=.52\linewidth]{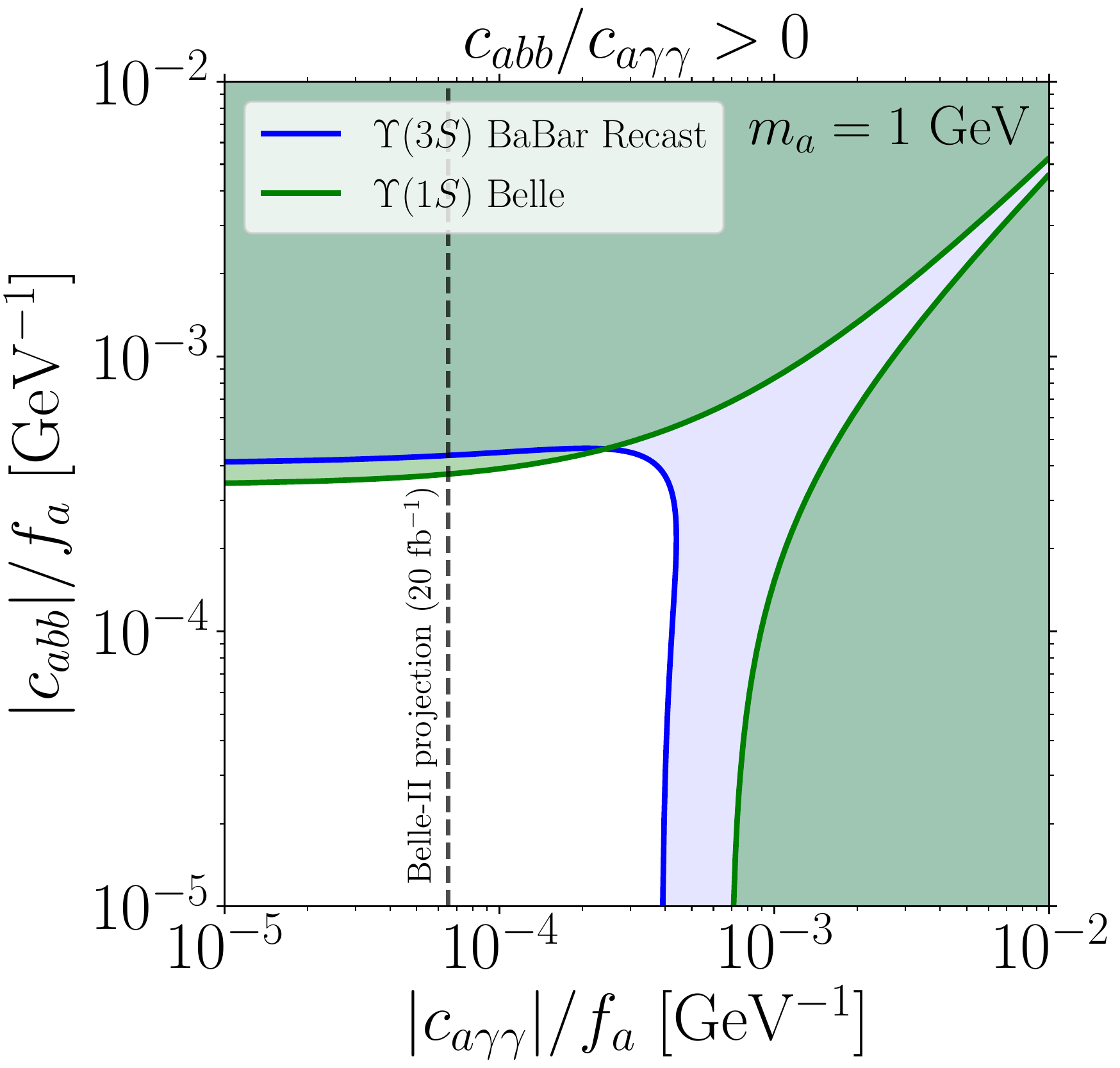}~\hspace*{-0.9em}
\includegraphics[width=.52\linewidth]{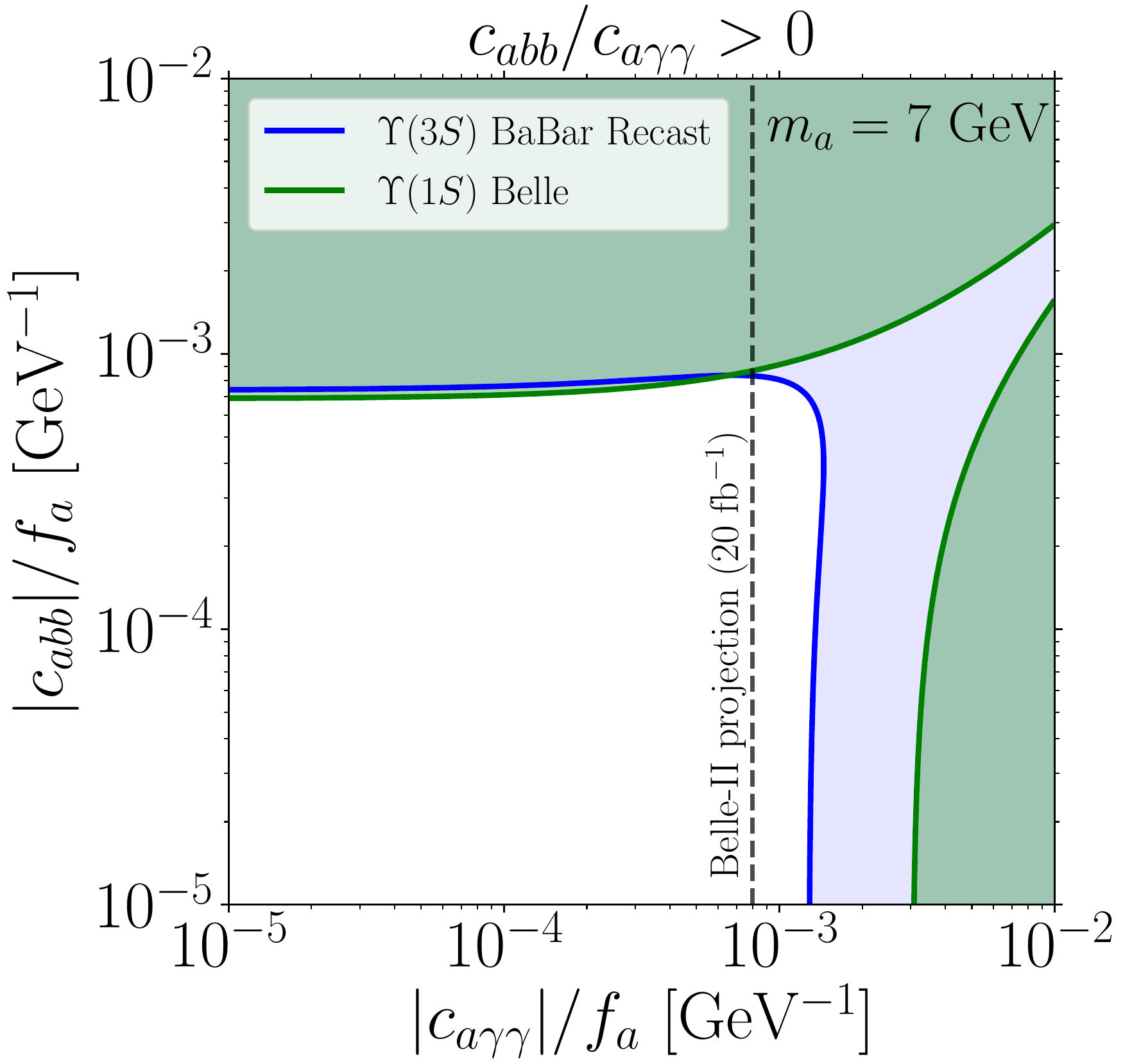}
\caption{\small \sl Excluded $\lbrace c_{a\gamma\gamma},c_{abb}\rbrace/f_a$ parameter space for the invisible ALP scenario when two couplings are simultaneously present. Belle constraints \cite{Seong:2018gut} at $\Upsilon(1S)$ (green line) and our recast of BaBar constraints \cite{Aubert:2008as} at $\Upsilon(3S)$ (blue line) are superimposed for the illustrative cases with $m_a = 1$~GeV (left panel) and $m_a = 7$~GeV (right panel). Projections for Belle-II sensitivity are depicted by the dashed lines. See text for details.}
\label{fig:constraints-alps-2param}
\end{figure*}

Next, the allowed parameter space in the plane $\lbrace c_{a\gamma\gamma},c_{abb}\rbrace/f_a$ when both couplings are simultaneously considered is shown in Fig.~\ref{fig:constraints-alps-2param}. To this purpose, two fixed values of $m_a$ are taken, namely $1$~GeV (left panel) and $7$~GeV (right panel), and $c_{abb}/c_{a\gamma\gamma}>0$ is assumed, in such a way that both couplings interfere destructively in Eq.~\eqref{eq:ups-general}. In this case, it can be seen from Fig.~\ref{fig:constraints-alps-2param} that the $\Upsilon(1S)$ constraints have an unconstrained direction that cannot be resolved by only relying on resonant ALP searches.~\footnote{A similar observation has been recently made for ALP produced in the rare decays $K^+\to \pi^+ a$ and $B \to K^{(\ast)} a$, for which the top-quark and $W$ loops can interfere destructively~\cite{Gavela:2019wzg}.} The combination of couplings that lead to this cancellation depends on the ALP mass, especially for $m_a$ values near the kinematical threshold, as depicted in the right panel of Fig.~\ref{fig:constraints-alps-2param}. BaBar results obtained at the $\Upsilon(3S)$ resonance, which is not reconstructed, depicts a different sensitivity to $\lbrace c_{a\gamma\gamma},c_{abb}\rbrace$, as shown by the blue regions in the same plot.  While a cancellation between $c_{a\gamma\gamma}$ and $c_{abb}$ is possible for resonant cross-section, this cannot occur for the non-resonant one \eqref{eq:sigma-monogamma}, which depends only on the $c_{a\gamma\gamma}$ coupling. The combination of these complementary searches allows one to corner the ALP parameter space as depicted in Fig.~\ref{fig:constraints-alps-2param}. Moreover, projections for searches performed at Belle-II, operating at the $\Upsilon(4S)$ resonance, as computed in Ref.~\cite{Dolan:2017osp}, are displayed in the same plot for an expected luminosity of $20~\mathrm{fb}^{-1}$.

Before concluding, comments on studies providing similar constraints on ALP couplings are needed. The authors of Ref.~\cite{Dolan:2017osp} have performed a reinterpretation of the BaBar dark-photon search in the $e^+e^-\to \gamma + \mathrm{inv}$ channel~\cite{Lees:2017lec}. The constraints on $c_{a\gamma\gamma}$ they obtain, by only considering the non-resonant process from Fig.~\ref{fig:diag-non-resonant}, are a factor of $\approx 2$ better than the limits derived in this paper. Nonetheless, such reinterpretation should be performed with caution for two reasons. Firstly, the kinematical distribution of this process is different for ALPs and dark photons scenarios, as can be inferred from the comparison between Eq.~\eqref{eq:sigma-monogamma} with the expressions given in Ref.~\cite{Essig:2009nc}. Therefore, to translate the dark photon constraints into limits on ALP couplings,  one should correct for the different detector efficiencies for the two cases. Another important issue is 
the fact that the dark photon analysis from Ref.~\cite{Lees:2017lec} combine off-resonance data with data collected at the $\Upsilon(2S)$, $\Upsilon(3S)$ and $\Upsilon(4S)$ resonances. While the photons accompanied by dark photons cannot be produced via $\Upsilon(nS)$ decays, this is not the case for ALPs, as discussed above. Therefore, it is important to account for the resonant ALP production estimated in Tab.~\ref{tab:xsection}, which is different for each data set considered by BaBar and which can amount to corrections of $\mathcal{O}(50\%)$ to the total cross-section.

\section{Conclusions}
\label{sec:summary}

In this paper, ALP production in association with photons at $B$-factories is revisited. In particular, the contributions to the $e^+ e^- \to \gamma a$ cross section are derived, assuming generic non-vanishing ALP couplings with both photons and $b$-quarks. The production of ALPs can proceed through the non-resonant channel, $e^+ e^- \to \gamma a$, as well as the resonant one, $e^+ e^- \to 
\Upsilon(nS) \to \gamma a$, which has the unique potential to probe the ALP coupling to $b$-quarks. After computing the relevant cross-sections and accounting for the effects stemming from the beam-energy uncertainty at $B$-factories, three distinct and complementary experimental searches have been identified: 
\begin{itemize} 
 \item[{i)}] \emph{Resonant searches} that exploit decays such as $\Upsilon(2S) \to \Upsilon(1S)\,\pi^+\pi^-$ and/or $\Upsilon(3S)\to \Upsilon(1S)\, 
  \pi^+\pi^-$ to directly probe the $\Upsilon(1S)$ decays~\cite{delAmoSanchez:2010ac,Seong:2018gut}, which turn out to be equally sensitive to ALP couplings to photons and 
  bottom quarks, as shown in Eq.~\eqref{eq:ups-general}; 
  \item[{ii)}]  \emph{Mixed (non-)resonant searches} that use, instead, the primarly produced $\Upsilon(nS)$ resonance, with $n=1,2,3$, as in the analysis performed in Ref.~\cite{Aubert:2008as}. These searches can probe both resonant and non-resonant ALP production, and hence are more sensitive to the ALP coupling to photons than to the one with $b$-quarks, cf.~Sec.~\ref{sec:exp}; 
  \item[{iii)}]  \emph{Non-resonant searches}, as the ones performed at $\sqrt{s}=m_{\Upsilon(4S)}$, that can provide information only on the ALP coupling to photons, cf.~Table~\ref{tab:xsection}. Note, in particular, that neither Babar or Belle have reported such an analysis thus far. 
\end{itemize}

\noindent Previous phenomenological analyses overlooked the distinction between these types of experimental searches, which have been clarified in this paper, and the optimal experimental strategies have also been discussed.

To illustrate the phenomenological implications of the effects mentioned above, the scenario with an ALP decaying into invisible final states has been considered. Constraints on the parameter space $\lbrace m_a; \,c_{a\gamma\gamma}, c_{abb} \rbrace$ have been derived from existing BaBar and Belle data, and projections for Belle-II have been discussed. In particular, constraints from resonant searches have a flat direction due to possible cancellations between $c_{a\gamma\gamma}$ and $c_{abb}$ contributions in $\mathcal{B}(\Upsilon (1S)\to \gamma a)$. These flat directions, however, can be removed by existing mixed (non-)resonant searches performed at $\sqrt{s}=m_{\Upsilon(3S)}$, due to the interplay between resonant and non-resonant contributions described above. In the future, the Belle-II experiment has the great opportunity to perform a first search at $\sqrt{s}=m_{\Upsilon(4S)}$, probing solely the $c_{a\gamma\gamma}$ coupling, and providing a complementary piece of information to the aforementioned constraints. Finally, the \textit{invisible} ALP scenario has been considered in this paper for sake of illustration, but the conclusions derived above also apply to scenarios where the ALP decays into visible particles. The phenomenological study of more general scenarios, including visible ALP decays, as well as experimental signatures with displaced vertices, will be the object of a future work.

\section{Acknowledgments}
\label{sec:acknowledgments}

The authors acknowledge F.~Anulli, D.~Becirevic, S.~Fajfer, A.~Guerrera, C.~Hearty, S.~J.D.~King, T.~Ferben, S.~Lacaprara, M.~Margoni, F.~Mescia, M.~Passera and P.~Paradisi for very useful exchanges. This project has received support by the European Union's Horizon 2020 research and innovation programme under the Marie Sklodowska-Curie grant agreement N$^\circ$~674896 (ITN Elusives) and 690575 (RISE InvisiblePlus) and by the exchange of researchers project ``The flavor of the invisible universe" funded by the Italian Ministry of Foreign Affairs and International Cooperation (MAECI). L.M. acknowledges partial financial support by the Spanish MINECO through the ``Ram\'on y Cajal'' programme (RYC-2015-17173), by the Spanish ``Agencia Estatal de Investigaci\'on'' (AEI) and the EU ``Fondo Europeo de Desarrollo Regional'' (FEDER) through the project FPA2016-78645-P, and through the Centro de excelencia Severo Ochoa Program under grant SEV-2016-0597. L.M. thanks the Physics and Astronomy Department 'G.Galilei' of the Universit\`a degli Studi di Padova for hospitality during the development of this project.

\end{document}